# Adaptive Monotone Shrinkage for Regression

**Zhuang Ma**
Department of Statistics
University of Pennsylvania
Philadelphia, PA 19104
zhuangma@wharton.upenn.edu

**Dean Foster**
Yahoo Lab
New York, NY 10018
dean@foster.net

**Robert Stine**
Department of Statistics
University of Pennsylvania
Philadelphia, PA 19104
stine@wharton.upenn.edu

## Abstract

We develop an adaptive monotone shrinkage estimator for regression models with the following characteristics: i) dense coefficients with small but important effects; ii) *a priori* ordering that indicates the probable predictive importance of the features. We capture both properties with an empirical Bayes estimator that shrinks coefficients monotonically with respect to their anticipated importance. This estimator can be rapidly computed using a version of Pool-Adjacent-Violators algorithm. We show that the proposed monotone shrinkage approach is competitive with the class of all Bayesian estimators that share the prior information. We further observe that the estimator also minimizes Stein's unbiased risk estimate. Along with our key result that the estimator mimics the oracle Bayes rule under an order assumption, we also prove that the estimator is robust. Even without the order assumption, our estimator mimics the best performance of a large family of estimators that includes the least squares estimator, constant-$\lambda$ ridge estimator, James-Stein estimator, etc. All the theoretical results are nonasymptotic. Simulation results and data analysis from a model for text processing are provided to support the theory.

## 1 INTRODUCTION

Feature selection and coefficient estimation are familiar topics in both statistics and machine learning communities. Many results in this area concern models that are 'nearly black,' possessing a handful of large effects against a wide field of noise. Consider the widely used linear model

$$Y = X\beta + \epsilon \quad \text{where} \quad \epsilon \sim N(0, \sigma^2 I_n), \quad (1)$$

$X$ is full-rank, $n \times p$ matrix of explanatory features with $p \leq n$, and $\beta$ is a $p$ dimensional vector of unknown coefficients. In the 'nearly black' case, all but a few of the coordinates of $\beta$ are zero. A long sequence of results leverage this sparsity (Foster and George [1994]; Tibshirani [1996]; Abramovich et al. [2006]; Candes and Tao [2007]; Fan and Lv [2008]; Bickel et al. [2009]). Sparsity assumptions are well suited to many applications, especially within the field of signal and image processing (Donoho [1995], Wright et al. [2009]).

Despite the prevalence of research on sparse models, some applications do not conform to this paradigm. For example, Foster et al. [2013] used methods such as latent semantic analysis, essentially principal components analysis (PCA), to convert text into features for regression analysis. The estimated coefficients of these principal components show two specific characteristics that draw our attention: dense coefficient estimates with a monotonically decaying effect size. Rather than concentrate in a few estimates, the predictive power of the model spreads across many features. Sparsity-based methods such as hard or soft thresholding that set small effects to zero produce fitted models with greatly diminished predictive ability. Too much predictive signal has been lost by eliminating small, but nonetheless informative, coefficients. Dense coefficients appear in other applications as well. Hall et al. [2009] and Dicker [2011, 2012] also propose models for dense signals and Dicker [2011, 2012] discusses several shrinkage estimators in high dimensions.

The second characteristic of this application is the monotone decrease in typical effect size. The signal tends to concentrate in the leading principal components, then gradually decay. We may not know the signal strength, but we do have an ordering. In this sense, the unsupervised PCA of the text data provides useful information that can be exploited within the regression. In particular, the eigenvalues from the PCA provide an external ordering of the features that is suggestive of the effect size. Such exogenous information appears in other domains. In time series analysis, data collected more recently are expected to be more informative for the prediction of future trends. In principal components regression, we tend to expect the first principal component to be more important than the second. Therefore, models without incorporating this prior knowledge might be suboptimal.

In this paper, we capture both characteristics with an empirical Bayes estimator that shrinks coefficients monotonically with respect to their anticipated importance. The procedure is tuning free and can be efficiently implemented using Pool-Adjacent-Violators algorithm. We further show that the estimator can be derived from frequentists' perspective as well by minimizing Stein's unbiased risk estimate. Finally, we establish non-asymptotic results to guarantee that the proposed estimator is nearly Bayes optimal under the order assumption and even when the order assumption (or say, prior knowledge) is wrong, it still mimics the best performance of a large family of estimators that includes the least squares estimator, ridge estimator, James-Stein estimator, etc.

The rest of this paper is organized as follows. In section 2, we describe the monotone shrinkage model in detail and introduce the maximum marginal likelihood estimator(MMLE) and Pool-Adjacent-Violators algorithm. In section 3, we show that the proposed estimator also minimizes Stein's unbiased risk estimate and establish its non-asymptotic oracle properties both with and without order assumption. In section 4, we suggest an estimator of the error variance $\sigma^2$. In section 5, we present simulation results and an analysis of text to support our theory. Concluding remarks are given in section 6. Details of the technique are provided in the Appendix.

## 2 ADAPTIVE MONOTONE SHRINAKGE

### 2.1 Model Formulation

We use a Bayesian framework to encode the prior knowledge about the importance of explanatory features in our model. We express this prior knowledge in a distribution of the coefficients $\beta$. Intuitively, if the features within the regression are standardized such that $X^T X = I_p$, then the coefficient $|\beta_i|$ gives the importance of the $i$th feature: a unit change in $X_i$ is associated with a change of $|\beta_i|$ in the response. A natural prior that captures the sense that the elements of $\beta$ have decaying size specifies a monotone decreasing sequence of variances for the coefficients. For convenience, we assume that the size of signal in $\beta_i$ is decreasing with the index $i$:

$$\beta_i \sim N(0, \sigma_i^2)$$
$$\sigma_1^2 \geq \sigma_2^2 \geq \cdots \geq \sigma_p^2 \geq 0 \qquad (2)$$

Since we know the order, we can always rearrange $\beta_i$ so that the unknown $\sigma_i^2$ are monotone as above. Throughout, we consider only orthonormal designs for which $X^T X = I_p$. Then the Bayes rule $\beta^*$ in (1) and (2) is:

$$\beta_i^* = \frac{\sigma_i^2}{\sigma_i^2 + \sigma^2}\tilde{\beta}_i \;,$$

where $\tilde{\beta} = (\tilde{\beta}_1, \cdots, \tilde{\beta}_p) = X^T Y$ is the least squares estimator. The Bayes rule shrinks $\tilde{\beta}_i$ monotonically, shrinking more and more harshly as the index and $\sigma_i^2$ increase. For our application, we know only the order of the features, not the signal strength $\sigma_i^2$, so the Bayes rule is not a real estimator because it depends on the unknown parameter $\sigma_i^2$. To mimic the performance of the Bayes rule, we estimate the $\sigma_i^2$s from data under the order constraint and use the resulting plug-in estimator. For convenience, we write the model as $\beta \sim N(0, \Sigma)$ where $\Sigma = diag(\sigma_1^2, \cdots, \sigma_p^2)$.

### 2.2 Maximum Marginal Likelihood Estimator

To estimate the ordered prior variances on the diagonal of $\Sigma$, we observe that the marginal distribution of $Y$ is $N(0, X\Sigma X^T + \sigma^2 I_n)$. If we further assume the error variance $\sigma^2$ from (1) is known (or we can plug in a consistent estimator), a simple calculation shows that the least square estimator $\tilde{\beta} = X^T Y$ is a sufficient statistic for $\Sigma$. So, in the following discussions, we base our inference on $\tilde{\beta}$, whose marginal distribution is $N(0, \sigma^2 I_p + \Sigma)$. A natural estimator of $\Sigma$ is the maximum marginal likelihood estimator (MMLE). The log marginal likelihood function is

$$l(\Sigma) = -\frac{1}{2}\sum_{i=1}^{p}\left(\log(2\pi) + \log(\sigma^2 + \sigma_i^2) + \frac{\tilde{\beta}_i^2}{\sigma^2 + \sigma_i^2}\right)$$

Consequently, the MMLE is the solution to the following optimization problem:

$$\arg\min_{\sigma_i} \sum_{i=1}^{p}\left(\log(\sigma^2 + \sigma_i^2) + \frac{\tilde{\beta}_i^2}{\sigma^2 + \sigma_i^2}\right) \qquad (3)$$
$$\text{subject to } \sigma_1^2 \geq \sigma_2^2 \geq \cdots \geq \sigma_p^2 \geq 0$$

### 2.3 Pool-Adjacent-Violators Algorithm

The optimization problem (3) resembles the well-known isotonic regression problem

$$\hat{\beta}^{iso} = \arg\min_{\beta} \sum_{i=1}^{n}(y_i - \beta_i)^2 \text{ subject to } \beta_1 \geq \cdots \geq \beta_n \qquad (4)$$

whose unique solution can be efficiently obtained by running the Pool-Adjacent-Violators (PAV) algorithm. Roughly speaking, this algorithm solves (4) as follows. Set $i = 1$. Move to the right (increase the index $i$) until finding a pair $(y_i, y_{i+1})$ that violates the monotonicity constraint, that is $y_i < y_{i+1}$. Pool $y_i$ and the adjacent $y_{i+1}$ and replace both by their average. Next check whether $y_{i-1} < \frac{y_i + y_{i+1}}{2}$. If so, replace $(y_{i-1}, y_i, y_{i+1})$ with their average. Continue to the left until monotonicity is satisfied and then proceed to the right until the whole sequence is monotone. Hence, PAV algorithm outputs an decreasing blockwise constant sequence. As far as we know, the PAV algorithm dates back to Ayer et al. [1955], where it is used to compute the MLE of independent binomial distributions. Brunk [1955, 1958] considered rather general scenarios and established some consistency properties. According to Grotzinger and Witzgall [1984], if carefully implemented, the PAV algorithm has computational complexity $O(n)$.

Although the optimization problem (3) is not convex, it can be solved efficiently by the PAV algorithm. Before establishing this result, we introduce some notations.

$$f_i(x) = \log(x + \sigma^2) + \frac{\tilde{\beta}_i^2}{x + \sigma^2}$$
$$\tilde{\sigma}_i^2 = \arg\min_x f_i(x) = \tilde{\beta}_i^2 - \sigma^2$$

**Proposition 2.1.** *The following two-step algorithm produces the MMLE denoted by $(\hat{\sigma}_1^2, \cdots, \hat{\sigma}_p^2)$*
*Step 1.* $(\check{\sigma}_1^2, \cdots, \check{\sigma}_p^2) = PAV(\tilde{\sigma}_1^2, \cdots, \tilde{\sigma}_p^2)$
*Step 2.* $\hat{\sigma}_i^2 = \check{\sigma}_i^2 I_{(\check{\sigma}_i^2 \geq 0)}$.

For those $\sigma_i^2$ estimated to be 0, it means the corresponding features are not included in the model. To prove the proposition, we first introduce a lemma:

**Lemma 2.1.** *Consider optimization problem*

$$\min \sum_{i=1}^p f_i(\theta_i)$$
*subject to:* $\theta_1 \geq \theta_2 \geq \cdots \geq \theta_p$

*where the elementwise solution $\tilde{\theta}_i = \arg\min_\theta f_i(\theta)$ is finite. If $\{f_1, \cdots, f_p\}$ satisfies the pooling condtion defined below, then the optimization problem has a unique solution $(\hat{\theta}_1, \cdots, \hat{\theta}_p)$=$PAV(\tilde{\theta}_1, \cdots, \tilde{\theta}_p)$.*

**Definition 1** (Pooling Condition). *For a sequence of functions $\{f_1, \cdots, f_p\}$ with $\tilde{\theta}_i = \arg\min_\theta f_i(\theta)$ being finite. Let $\overline{\theta}_{ij} = \sum_{k=i}^j \tilde{\theta}_k/(j-i+1)$. We say $\{f_1, \cdots, f_p\}$ satisfies pooling condition if $\forall i \leq j, \arg\min_\theta \sum_{k=i}^j f_k(\theta) = \overline{\theta}_{ij}$ and $\sum_{k=i}^j f_k(\theta)$ is strictly decreasing when $\theta \leq \overline{\theta}_{ij}$ and strictly increasing when $\theta \geq \overline{\theta}_{ij}$.*

The pooling condition can be easily checked and numerous distributions have log likelihood functions that satisfy the condition. These include the binomial distribution, Poisson distribution, normal distribution with fixed variance and variable mean, normal distribution with fixed mean and variable variance, and so on.

*Proof of Proposition 2.1*
The situation we are faced up with is normal distribution with fixed mean and variable variance, which satisfies the conditions in Lemma 1. According to the lemma,

$(\check{\sigma}_1^2, \cdots, \check{\sigma}_p^2) = PAV(\tilde{\sigma}_1^2, \cdots, \tilde{\sigma}_p^2)$ solves:

$$\min \sum_{i=1}^p (\log(\sigma^2 + \sigma_i^2) + \frac{\tilde{\beta}_i^2}{\sigma^2 + \sigma_i^2})$$
subject to: $\sigma_1^2 \geq \sigma_2^2 \geq \cdots \geq \sigma_p^2$

which is only slightly different from our original optimization problem. To finish the proof, we just need to introduce some auxiliary functions. Let $f_{-k} = \log(\sigma^2 + \sigma_{-k}^2) + \frac{\sigma^2}{\sigma^2 + \sigma_{-k}^2}, 1 \leq k \leq n$, so $\tilde{\sigma}_{-k}^2 = \arg\min f_{-k}(\sigma_{-k}^2) = 0$. Consider the following optimization problem $(Q^n)$:

$$\min \sum_{i=1}^p (\log(\sigma^2 + \sigma_i^2) + \frac{\tilde{\beta}_i^2}{\sigma^2 + \sigma_i^2})$$
$$+ \sum_{i=1}^n (\log(\sigma^2 + \sigma_{-i}^2) + \frac{\sigma^2}{\sigma^2 + \sigma_{-i}^2})$$
subject to $\sigma_1^2 \geq \cdots \geq \sigma_p^2 \geq \sigma_{-1}^2 \geq \cdots \geq \sigma_{-n}^2$

Lemma 1 shows $PAV(\tilde{\sigma}_1^2, \cdots, \tilde{\sigma}_p^2, 0, \cdots, 0)$ solves $(Q^n)$. Denote it $(\hat{\sigma}_{n1}^2, \cdots, \hat{\sigma}_{np}^2, \hat{\sigma}_{-nn}^2, \cdots, \hat{\sigma}_{-n1}^2)$. Notice that $(\hat{\sigma}_1^2, \cdots, \hat{\sigma}_p^2, 0, \cdots, 0)$ is a feasible solution, which should be suboptimal, that is to say,

$$\sum_{i=1}^p (\log(\sigma^2 + \hat{\sigma}_{ni}^2) + \frac{\tilde{\beta}_i^2}{\sigma^2 + \hat{\sigma}_{ni}^2}) + \sum_{i=1}^n f_{-k}(\hat{\sigma}_{-ni}^2)$$
$$\leq \sum_{i=1}^p (\log(\sigma^2 + \hat{\sigma}_i^2) + \frac{\tilde{\beta}_i^2}{\sigma^2 + \hat{\sigma}_i^2}) + \sum_{i=1}^n f_{-k}(0)$$

Recall that $\tilde{\sigma}_{-k}^2 = \arg\min f_{-k}(\sigma_{-k}^2) = 0$, which implies

$$\sum_{i=1}^p (\log(\sigma^2 + \hat{\sigma}_{ni}^2) + \frac{\tilde{\beta}_i^2}{\sigma^2 + \hat{\sigma}_{ni}^2})$$
$$\leq \sum_{i=1}^p (\log(\sigma^2 + \hat{\sigma}_i^2) + \frac{\tilde{\beta}_i^2}{\sigma^2 + \hat{\sigma}_i^2})$$

Let $n$ goes to infinity, $PAV(\tilde{\sigma}_1^2, \cdots, \tilde{\sigma}_p^2, 0, \cdots, 0)$ converges to $(\check{\sigma}_1^2 I_{(\check{\sigma}_1^2 \geq 0)}, \cdots, \check{\sigma}_p^2 I_{(\check{\sigma}_p^2 \geq 0)}, 0, \cdots, 0)$ and the inequality above implies $(\check{\sigma}_1^2 I_{(\check{\sigma}_1^2 \geq 0)}, \cdots, \check{\sigma}_p^2 I_{(\check{\sigma}_p^2 \geq 0)})$ is the solution to the original optimization problem. □

## 2.4 Data-Driven Blockwise James-Stein Estimator: Global and Local Adaptivity

Proposition 1 and the nature of Pool-Adjacent-Violators algorithm show that the MMLE $(\hat{\sigma}_1^2, \cdots, \hat{\sigma}_p^2)$ is decreasing and blockwise constant. We now change notation in this part and let $\hat{\sigma}_i^2$ denote the common variance estimate of the $i$th block. Define $\beta_i = (\beta_{i1}, \cdots, \beta_{in_i})$ to be the coefficients within the $i$th block and correspondingly define $\hat{\beta}_i, \tilde{\beta}_i$. With these notations, we can write explicitly

$$\hat{\sigma}_i^2 = \left(\frac{\sum_{j=1}^{n_i}(\tilde{\beta}_{ij}^2 - \sigma^2)}{n_i}\right)_+ \text{ and}$$

$$\begin{aligned}\hat{\beta}_i &= \frac{\hat{\sigma}_i^2}{\hat{\sigma}_i^2 + \sigma^2}\tilde{\beta}_i \\ &= (1 - \frac{n_i\sigma^2}{\sum_{j=1}^{n_i}\tilde{\beta}_{ij}^2})_+\tilde{\beta}_i,\end{aligned}$$

which is exactly the positive part of the James-Stein type estimator. Hence the proposed estimator can be interpreted as a monotone blockwise James-Stein estimator. Blockwise James-Stein estimator is well-studied in the wavelet setting (Cai [1999], Cai and Zhou [2009]). In Cai [1999], the block size is fixed before observing the data and is the same for all blocks. Cai and Zhou [2009] proposed an adaptive procedure to make the block size data-driven but the block size remains the same for all blocks. As for our procedure, the number of blocks and the size of each block are completely data-driven. The difference is due to different assumptions. The former is based on smoothness of Besov bodies while the later is based on monotonicity.

The advantage of our data-driven, monotone blockwise James-Stein estimator is the ability to achieve both global and local adaptivity. Blockwise shrinkage utilizes information about neighboring coefficients. However, if the block size is too large, local inhomogeneity might be overlooked. So, the best way to achieve a good balance is to let the data speak for itself.

## 3 ORACLE RISK PROPERTIES

### 3.1 Equivalence between MMLE and SURE Estimator

In this section, we show that our empirical Bayes estimator can also be derived within a frequentist framework by minimizing Stein's unbiased risk estimate (SURE). Under squared error loss: $l(\hat{\beta}, \beta) = \frac{1}{p}\sum_{i=1}^{p}(\hat{\beta}_i - \beta_i)^2$. If one uses the shrinkage estimator $\hat{\beta}^\lambda$ defined by $\hat{\beta}_i^\lambda = \frac{\lambda_i}{\lambda_i + \sigma^2}\tilde{\beta}_i$ to estimate $\beta_i$, the risk for a given $\beta$ is:

$$R_p(\hat{\beta}^\lambda, \beta) = E[l(\hat{\beta}^\lambda, \beta)] = \frac{1}{p}\sum_{i=1}^{p}\frac{\sigma^2}{(\sigma^2 + \lambda_i)^2}(\sigma^2\beta_i^2 + \lambda_i^2)$$

and an unbiased estimate for the risk is

$$SURE(\lambda) = \frac{1}{p}\sum_{i=1}^{p}[(\frac{\sigma^2}{\sigma^2 + \lambda_i})^2\tilde{\beta}_i^2 + \frac{\sigma^2(\lambda_i - \sigma^2)}{\sigma^2 + \lambda_i}]$$

Generally, $SURE(\lambda)$ is unbiased estimate of the risk only if $\lambda$ is a fixed constant and cannot depend on data. We say $\hat{\beta}^{\hat{\lambda}}$ is a **monotone shrinkage estimator** if $\hat{\beta}_i^{\hat{\lambda}} = \frac{\hat{\lambda}_i}{\hat{\lambda}_i + \sigma^2}\tilde{\beta}_i$ and $\hat{\lambda}_1 \geq \cdots \geq \hat{\lambda}_p \geq 0$, where $\hat{\lambda}_i$ can be data dependent. A monotone shrinkage estimator is completely determined by the **monotone shrinkage parameter** $\hat{\lambda} = (\hat{\lambda}_1, \cdots, \hat{\lambda}_p)$. If only considering the family of monotone shrinkage estimators, the relationship suggests that the data-dependent $\hat{\lambda}$ which minimizes SURE($\lambda$) should be a good choice. Define:

$$\hat{\lambda}_{SURE} = \arg\min_{\lambda_1 \geq \cdots \geq \lambda_p \geq 0}\sum_{i=1}^{n}[(\frac{\sigma^2}{\sigma^2 + \lambda_i})^2\tilde{\beta}_i^2 + \frac{\sigma^2(\lambda_i - \sigma^2)}{\sigma^2 + \lambda_i}]$$

which is of the same form as optimization problem (3). Let $g_i(\lambda_i) = (\frac{\sigma^2}{\sigma^2 + \lambda_i})^2\tilde{\beta}_i^2 + \frac{\sigma^2(\lambda_i - \sigma^2)}{\sigma^2 + \lambda_i}$. Then it is easy to see that $\tilde{\lambda}_i = \arg\min_{\lambda_i} g_i(\lambda_i) = \tilde{\beta}_i^2 - \sigma^2$. Checking that $g_i(\lambda_i)$ satisfy the two conditions in Lemma 2.1, the same argument used to show Proposition 2.1 implies:

**Proposition 3.1.** *MMLE equals SURE estimator $\hat{\beta}^{\hat{\lambda}_{SURE}}$.*

In the rest of the paper, we will use $\hat{\beta}_{SURE} = \hat{\beta}^{\hat{\lambda}_{SURE}}$ to denote the proposed estimator.

**Remark 3.1.** *Monotone shrinkage estimator was also investigated in Xie et al. [2012] when dealing with heteroscedastic normal sequence model. Different empirical Bayes estimators were studied in this paper and SURE estimator was shown to dominate MMLE and method of moments. While in our context, the three estimators turned out to be the same.*

### 3.2 Oracle Property with Order Assumption

Proposition 3.1 provides us with a powerful tool to investigate the risk properties of the proposed estimate. First of all, we introduce the oracle estimator, namely the Bayes rule $\beta^* = (\beta_1^*, \cdots, \beta_p^*)$ defined by

$$\beta_i^* = \frac{\sigma_i^2}{\sigma_i^2 + \sigma^2}\tilde{\beta}_i$$

Of course, $\beta^*$ is not an practical estimator because it depends on the unknown parameter $\lambda^* = (\sigma_1^2, \cdots, \sigma_p^2)$. It is easy to see the oracle risk is:

$$R(\beta^*) = \frac{1}{p}\sum_{i=1}^{p}\frac{\sigma^2\sigma_i^2}{\sigma^2 + \sigma_i^2}$$

Then we introduce another lemma, which is the building block of the oracle properties. It says that $E[SURE(\hat{\lambda})]$ is uniformly good approximation of the true risk $E[l(\hat{\beta}_{\hat{\lambda}})]$, where the expectation is with respect to both data and parameter $\beta$.

**Lemma 3.1.**

$$\sup_{\sigma_1^2, \cdots, \sigma_p^2}\sup_{\hat{\lambda}_1 \geq \cdots \geq \hat{\lambda}_p}|E\{E[l(\hat{\beta}_{\hat{\lambda}}, \beta) - SURE(\hat{\lambda})|\beta]\}| \leq 4\sqrt{\frac{2}{p}}\sigma^2$$

*where $\hat{\lambda} = (\hat{\lambda}_1, \cdots, \hat{\lambda}_p)$ is arbitrary monotone shrinkage parameter and can be data dependent.*

**Theorem 3.1.**
$$\sup_{\sigma_1^2 \geq \cdots \geq \sigma_p^2 \geq 0} (R(\hat{\beta}_{SURE}) - R(\beta^*)) \leq 4\sqrt{\frac{2}{p}}\sigma^2$$

*Proof:* Because $\lambda^*$ is fixed constant, we have

$$R(\hat{\beta}_{SURE}, \beta) - R(\beta^*, \beta)$$
$$= E[l(\hat{\beta}_{SURE}, \beta)|\beta] - E[SURE(\lambda^*)|\beta]$$
$$= E[l(\hat{\beta}_{SURE}, \beta) - SURE(\hat{\lambda}_{SURE}) +$$
$$SURE(\hat{\lambda}_{SURE}) - SURE(\lambda^*)|\beta]$$
$$\leq E[l(\hat{\beta}_{SURE}, \beta) - SURE(\hat{\lambda}_{SURE})|\beta]$$

The inequality is due to the definition of $\hat{\lambda}_{SURE}$. So the Bayes risk satisfies:

$$R(\hat{\beta}_{SURE}) - R(\beta^*)$$
$$\leq E\{E[l(\hat{\beta}_{SURE}, \beta) - SURE(\hat{\lambda}_{SURE})|\beta]\}$$
$$\leq \sup_{\sigma_1^2, \cdots, \sigma_p^2} \sup_{\hat{\lambda}_1 \geq \cdots \geq \hat{\lambda}_p \geq 0} |E\{E[l(\beta_{\hat{\lambda}}, \beta) - SURE(\hat{\lambda})|\beta]\}|$$

Applying lemma 3.1 finishes the proof. □

**Remark 3.2.** *Theorem 3.1 shows that SURE estimator mimics the oracle Bayes rule and therefore outperforms all other estimators. What needs to be highlighted is that this is a non-asymptotic result with rate of convergence $O(p^{-\frac{1}{2}})$ independent of the true $\sigma_i^2$s. No matter how the $\sigma_i^2$s vary, as long as the order is known, the proposed adaptive procedure can uniformly capture the truth.*

**Corollary 3.1.** $\sup_{\sigma_1^2 \geq \cdots \geq \sigma_p^2 \geq 0} \frac{R(\hat{\beta}_{SURE})}{\sigma^2 + R(\beta^*)} = 1 + 4\sqrt{\frac{2}{p}}$

### 3.3 Oracle Property without Order Assumption

In this section, we show that even without knowing the order of the $\sigma_i^2$'s, the proposed estimator retains an oracle property among monotone shrinkage estimators.

**Theorem 3.2.**
$$\sup_{\sigma_1^2, \cdots, \sigma_p^2} (R(\hat{\beta}_{SURE}) - \inf_{\hat{\gamma}_1 \geq \cdots \geq \hat{\gamma}_p \geq 0} R(\hat{\beta}_{\hat{\gamma}})) \leq 8\sqrt{\frac{2}{p}}\sigma^2$$

*Proof:* For any given $(\sigma_1^2, \cdots, \sigma_p^2)$, we can always find $\hat{\eta} = (\hat{\eta}_1, \cdots, \hat{\eta}_p)$ that satisfies $\hat{\eta}_1 \geq \cdots \geq \hat{\eta}_p \geq 0$ and $R(\hat{\beta}_{\hat{\eta}}) < \inf_{\hat{\gamma}_1 \geq \cdots \geq \hat{\gamma}_p \geq 0} R(\hat{\beta}_{\hat{\gamma}}) + \epsilon$. Then,

$$R(\hat{\beta}_{SURE}) - \inf_{\hat{\gamma}_1 \geq \cdots \geq \hat{\gamma}_p \geq 0} R(\hat{\beta}_{\hat{\gamma}}) \leq R(\hat{\beta}_{SURE}) - R(\hat{\beta}_{\hat{\eta}}) + \epsilon$$

Notice that,

$$l(\beta, \hat{\beta}_{SURE}) - l(\beta, \hat{\beta}_{\hat{\eta}}) = (l(\beta, \hat{\beta}_{SURE}) - SURE(\hat{\lambda}_{SURE}))$$
$$+ (SURE(\hat{\lambda}_{SURE}) - SURE(\hat{\eta})) + (SURE(\hat{\eta}) - l(\beta, \hat{\beta}_{\hat{\eta}}))$$
$$\leq (l(\beta, \hat{\beta}_{SURE}) - SURE(\hat{\lambda}_{SURE})) + (SURE(\hat{\eta}) - l(\beta, \hat{\beta}_{\hat{\eta}}))$$

Take expectations, we have

$$R(\hat{\beta}_{SURE}) - R(\hat{\beta}_{\hat{\eta}}) \leq E\{l(\beta, \hat{\beta}_{SURE}) -$$
$$SURE(\hat{\lambda}_{SURE}) + SURE(\hat{\eta}) - l(\beta, \hat{\beta}_{\hat{\eta}})|\beta]\}$$
$$\leq 2 \sup_{\sigma_1^2, \cdots, \sigma_p^2} \sup_{\hat{\lambda}_1 \geq \cdots \geq \hat{\lambda}_p \geq 0} |E\{E[l(\beta_{\hat{\lambda}}, \beta) - SURE(\hat{\lambda})|\beta]\}|$$

Lemma 3.1 implies,

$$R(\hat{\beta}_{SURE}) - \inf_{\hat{\gamma}_1 \geq \cdots \geq \hat{\gamma}_p \geq 0} R(\hat{\beta}_{\hat{\gamma}}) \leq 8\sqrt{\frac{2}{p}}\sigma^2 + \epsilon$$

Since the upper bound does not depend on $\sigma_i^2$, let $\epsilon \to 0$, and the theorem follows. □

If we replace the data dependent shrinkage parameters in Theorem 2 with fixed ones, we can improve the error bound by a factor of 2, which is

**Corollary 3.2.**
$$\sup_{\sigma_1^2, \cdots, \sigma_p^2} (R(\hat{\beta}_{SURE}) - \inf_{\gamma_1 \geq \cdots \geq \gamma_p \geq 0} R(\hat{\beta}_\gamma)) \leq 4\sqrt{\frac{2}{p}}\sigma^2$$

Theorem 2 shows that even when the order assumption is invalid, the proposed estimator is nearly the best in the family of monotone shrinkage estimators. In particular, uniform shrinkage estimators such as least square estimator, ridge estimator and James-Stein estimator and stepwise regression methods such as monotone AIC, BIC, RIC (just search for $p$ nested submodels: with $i$th submodel as $\{1, \cdots, i\}$) are included. This is also a non-asymptotic result with rate of convergence $O(p^{-\frac{1}{2}})$ independent of the $\sigma_i^2$s. Therefore, the proposed estimator is robust and good enough for practical use. Actually, Theorem 3.2 states about the worst case. If the order is partially right, the proposed procedure benefits where the order is right and retains good properties where the order is wrong.

**Remark 3.3.** *The robustness is due to the 'soft constraint'. Instead of restricting the norm of regression coefficients to be monotone, we incorporate the constraint in the prior distribution, which makes the model flexible and robust.*

## 4 ESTIMATION OF $\sigma^2$

We have assumed $\sigma^2$ is known to establish the theoretical properties of our estimator. Here we suggest a reasonable estimate in practice that is based on maximum marginal likelihood. Unlike section 2.2, within this section the unknown parameter becomes $\theta = (\sigma_1^2, \cdots, \sigma_p^2, \sigma^2)$. Recall that the marginal distribution of $Y$ is $N(0, X\Sigma X^T + \sigma^2 I_n)$. So, the log marginal likelihood function:

$$l(\theta|y) \propto -log(|X\Sigma X^T + \sigma^2 I_n|) - y^T(X\Sigma X^T + \sigma^2 I_n)^{-1}y$$

where $|\cdot|$ means determinant. Let $X = (x_1, \cdots, x_p)$, we can add another $n-p$ vectors $x_{p+1}, \cdots, x_n$ to make

$\tilde{X} = (X, x_{p+1}, \cdots, x_n)$ an orthonormal matrix. Let $\tilde{\Sigma} = diag(\Sigma + \sigma^2 I_p, \sigma^2 I_{n-p})$. Then $X\Sigma X^T + \sigma^2 I_n = \tilde{X}\tilde{\Sigma}\tilde{X}^T$ and thus $(X\Sigma X^T + \sigma^2 I_n)^{-1} = \tilde{X}\tilde{\Sigma}^{-1}\tilde{X}^T$. Plug this expression back into the marginal likelihood function,

$$l(\theta|y) \propto -log(|\tilde{X}\tilde{\Sigma}\tilde{X}^T|) - y^T \tilde{X}\tilde{\Sigma}^{-1}\tilde{X}^T y$$

We abuse notation in this section and let $\tilde{\beta} = \tilde{X}^T y$. If we introduce variable $(\tau_1^2, \cdots, \tau_n^2) = diag(\tilde{\Sigma}) = (\sigma_1^2 + \sigma^2, \cdots, \sigma_p^2 + \sigma^2, \sigma^2, \cdots, \sigma^2)$, then

$$l(\Sigma) \propto -\sum_{i=1}^{n}(\log \tau_i^2 + \frac{\tilde{\beta}_i^2}{\tau_i^2})$$

So, the MMLE is the solution to the following optimization problem.

$$\min \sum_{i=1}^{p}(\log \tau_i^2 + \frac{\tilde{\beta}_i^2}{\tau_i^2})$$
subject to: $\tau_1^2 \geq \cdots \geq \tau_p^2 \geq \tau_{p+1}^2 = \cdots = \tau_n^2 \geq 0$

Following a similar but slightly different argument in Proposition 2.1, we obtain the following result and omit the proof.

**Proposition 4.1.** *The solution is uniquely given by*

$$(\hat{\tau}_1^2, \cdots, \hat{\tau}_p^2, \hat{\tau}_{p+1}^2, \cdots, \hat{\tau}_n^2) = PAV(\tilde{\beta}_1^2, \cdots, \tilde{\beta}_p^2, \frac{\sum_{i=p+1}^n \tilde{\beta}_i^2}{n-p}, \cdots, \frac{\sum_{i=p+1}^n \tilde{\beta}_i^2}{n-p})$$

*The MMLE of original parameters can be recovered by* $(\hat{\tau}_1^2, \cdots, \hat{\tau}_p^2, \hat{\tau}_{p+1}^2, \cdots, \hat{\tau}_n^2) = (\hat{\sigma}_1^2 + \hat{\sigma}^2, \cdots, \hat{\sigma}_p^2 + \hat{\sigma}^2, \hat{\sigma}^2, \cdots, \hat{\sigma}^2)$.

## 5 NUMERICAL EXPERIMENTS

### 5.1 Simulation Results

In this section, we compare the proposed monotone shrinkage approach with several other popular methods for feature selection and estimation. For simplicity, we only consider the normal sequence model and assume the error variance $\sigma^2$ is known.

- PAV, the proposed adaptive monotone shrinkage procedure computed by Pool-Adjacent-Violators algorithm.
- Lasso, with $\lambda$ selected by minimizing Stein's unbiased risk estimate. Under orthogonal design, it is also known as Sureshrink (Donoho and Johnstone [1995]).
- Ridge estimator with $\lambda$ selected by Cross-Validation.
- Positive part of James-Stein estimator
- Classical stepwise regression, we use AIC for penalty criterion.
- Monotone AIC: AIC that just searches for $p$ nested submodels, i.e., with $k_{th}$ submodel=$\{1, \cdots, k\}$

We consider the following scenarios ($p = 100, \sigma^2 = 1$):

1. Signals with Decaying Size: $(\sigma_1^2, \cdots, \sigma_p^2)$ are generated from decreasing order statistics of $2\chi^2$.
2. Signals with Same Size: $\sigma_i^2 = 2, \forall 1 \leq i \leq p$
3. Sparse Signals: first 90% of the $\sigma_i^2$ are 0 and remaining 10% of $\sigma_i^2$ are generated from decreasing order statistics of $4\chi^2$.
4. Signals with Increasing Size: $(\sigma_1^2, \cdots, \sigma_p^2)$ are generated from increasing order statistics of $2\chi^2$. This scenario dose not satisfy our order assumption(actually, the worst case), which is used to show the robustness of our procedure.

With $(\sigma_1^2, \cdots, \sigma_p^2)$ fixed, we adopt the following simulation strategy.

1. Generate $\beta = (\beta_1, \cdots, \beta_p)$ by $\beta_i \sim N(0, \sigma_i^2)$
2. Condition on $\beta$, generate the observation $X = (x_1, \cdots, x_p)$ by $x_i \sim N(\beta_i, \sigma^2)$
3. Use the methods discussed above to estimate the signal $\beta$ and compute the mean square error.
4. Repeat 1-3 for 400 times. The average of the mean square errors is an estimate of the Bayes risk.

Mean square error for different $\beta$ are given below using box plot and the middle line of each box represents Bayes risk of each procedure. The **red line** in the figure stands for the oracle risk, i.e., the Bayes risk of the oracle Bayes rule.

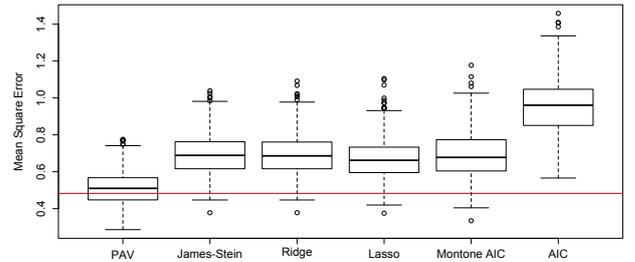

Figure 1: Signals with decaying size, i.e. $\{\sigma_i^2\}$ is decreasing. The adaptive monotone shrinkage procedure pools signals of similar sizes together and shrinks blockwisely and monotonically.

For signals with decaying size, oracle estimator shrink monotonically with respect to the size of the signals. Uniform shrinkage estimators such as ridge and James-Stein estimator are suboptimal. The proposed adaptive monotone procedure makes use of the prior information and mimics the oracle Bayes rule by pooling signals of similar size together so that it shrinks blockwisely and monotonically. AIC overfits the data while monotone AIC makes use of the order structure and therefore performs better.

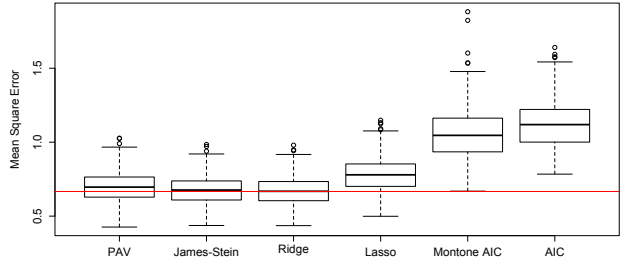

Figure 2: Signals with same size, i.e. $\sigma_i^2$s are the same. This is the case where ridge and James-Stein estimator capture the truth with full power while our procedure will regard the $\sigma_i^2$ as different(decreasing) and will generally divide the $\sigma_i^2$s into more than one blocks, which leads to slight power loss.

For signals with same size, the oracle estimator shrink uniformly. Ridge and James-Stein estimator mimic the oracle Bayes rule with full power. The proposed adaptive procedure does not necessarily gaurantee uniform shrinkage but the power loss is negligible.

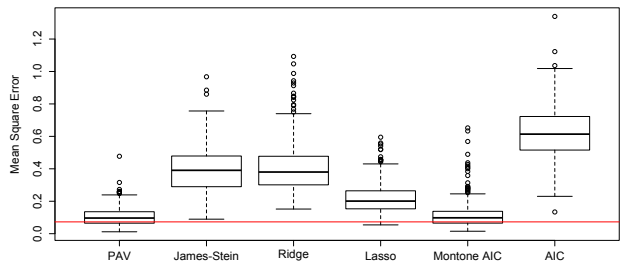

Figure 3: Sparse Signals, i.e. the size of the signals remain decreasing while 90% of them are 0. The proposed monotone shrinkage procedure can effectively kill the noise and shrink the signals properly.

For sparse signals, the oracle estimator kill the noise and shrink the signals monotonically. Monotone AIC can efficiently distinguish signal and noise while does not shrink the signals. The proposed adaptive procedure not only kills the noise but also shrinks the real signals properly according to their sizes. For those methods that cannot make use of the order structure, Lasso does better in this sparse case.

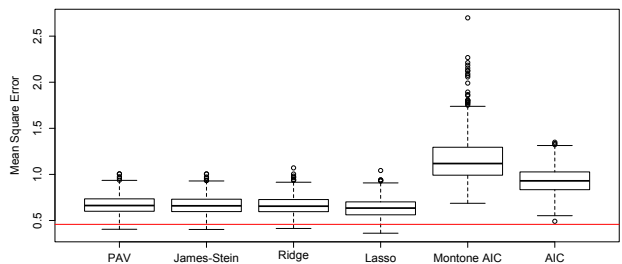

Figure 4: Signals with increasing size, i.e. $\{\sigma_i^2\}$ is increasing, which is opposite to our assumption that the size of signals is decaying. The adaptive monotone shrinkage procedure is robust and performs as good as other estimators.

For signals with increasing size, the proposed estimator uses completely reverse order. As theorem 2 expects, wrong prior knowledge won't ruin our estimator. It still mimics the best performance of the monotone shrinkage family. However, monotone AIC, which is not as robust as our procedure, suffers a lot from wrong prior knowledge.

### 5.2 Analysis of Text Processing Data

In this section, we apply the proposed adaptive monotone shrinkage approach to text data of real estate described in Foster et al. [2013]. The features included in the regression model are the leading 1500 principal components of the bag-of-words of text. The response is the log transformation of the real estate price. We use the eigenvalues from PCA to order the effect size of the features (see Figure 5 for the absolute t-statistics of the leading 500 principal components). Although the data dose not ideally satisfy the assumptions of our model, the proposed adaptive procedure is robust enough to leverage this rough prior knowledge.

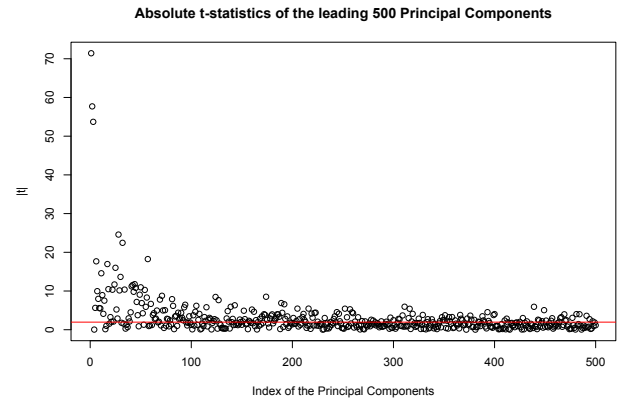

Figure 5: Absolute t-statistics of the leading 500 Principal Components. Those above the red line are significant.

The sample size is 7384 and we use 10 fold cross validation to estimate the prediction error of each procedure.

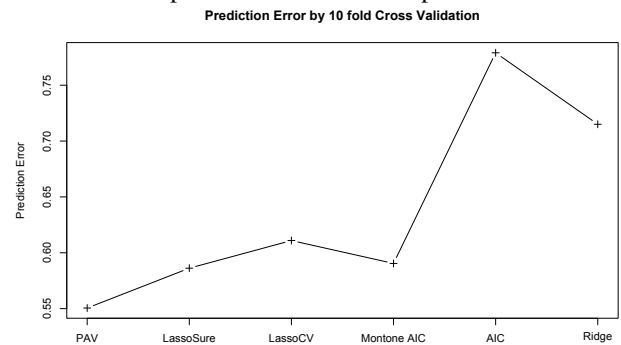

Figure 6: Prediction error comparison of different methods. LassoSURE: Lasso with tunning paramter selected by minimizing Stein's unbiased risk estimate. LassoCV: Lasso computed by LARS (Efron et al. [2004]) and paramter tuned by cross validation.

The result shows that

- PAV outperforms Ridge regression. From Figure 5, we can see that the signals are of different sizes. Uniform shrinkage method shrink the important features too much while shrink weak signals less harshly than it should be. PAV can adaptively pool the signals of similar size together and shrink blockwisely and monotonically.

- PAV outperforms LassoSure and LassoCV. Lasso can capture the sparse pattern of the data but as sacrifice, it might shrink important features a bit more than they should be.

- PAV outperforms Monotone AIC. Both procedures make use of prior information but PAV is more robust. There are several informative principal components corresponding to small eigenvalues so that Monotone AIC will exclude them from the model.

## 6 CONCLUSIONS

In this paper, we proposed an adaptive monotone shrinkage approach for regression with features of ordered effect size. We showed that the procedure can be rapidly computed via Pool-Adjacent-Violators algorithm and holds oracle risk properties. Non-asymptotic results are established. Furthermore, although the procedure is based on knowing the right prior knowledge about the features, we proved that, when the prior knowledge is wrong or in the absence of prior knowledge, the estimator still mimics the best performance of the family of monotone shrinkage estimators. Hence, it is robust enough to use in practice.

Compared with penalized least square methods which require heavy computational effort to find the best regularization paramter, the proposed adaptive procedure is tuning free. As noticed in the analysis of text data, the monotone shrinkage approach naturally works with PCA since the principal components are essentially ordered and orthogonal. Recent devolopments in randomized algorithms(Halko et al. [2011]) enable us to quickly compute the PCA of a huge matrix so that the proposed procedure can be easily applied to large-scale datasets.

## 7 APPENDIX

### 7.1 Proof of Lemma 2.1

It is sufficient to prove the following two claims:

i) $(\hat{\theta}_1^k, \cdots, \hat{\theta}_k^k)$=PAV$(\tilde{\theta}_1, \cdots, \tilde{\theta}_k), 1 \leq k \leq p$
ii) $\forall 1 \leq i, j \leq k, \hat{\theta}_i^k = \hat{\theta}_j^k \Rightarrow \hat{\theta}_i^m = \hat{\theta}_j^m, \forall k \leq m \leq p$

We prove the claim by induction:

1. It is trivial for $k = 1$ since $\hat{\theta}_1^1 = \tilde{\theta}_1$

2. Assuming the claim holds for $k$ and we are going to prove it is valid for $k + 1$ as well.

**Case 1** $\tilde{\theta}_{k+1} < \hat{\theta}_k^k$. It implies that

$$\sum_{i=1}^{k} f_i(\hat{\theta}_i^k) + f_{k+1}(\tilde{\theta}_{k+1})$$
$$= \min_{\theta_1 \geq \cdots \geq \theta_k} \sum_{i=1}^{k} f_i(\theta_i) + \min_{\theta_{k+1}} f_{k+1}(\theta_{k+1})$$
$$\leq \min_{\theta_1 \geq \cdots \geq \theta_{k+1}} \sum_{i=1}^{k+1} f_i(\theta_i)$$

and therefore,

$$(\hat{\theta}_1^{k+1}, \cdots, \hat{\theta}_{k+1}^{k+1}) = (\hat{\theta}_1^k, \cdots, \hat{\theta}_k^k, \tilde{\theta}_{k+1})$$
$$= PAV(\tilde{\theta}_1, \cdots, \tilde{\theta}_k)$$

So we prove claim i). Notice that $\hat{\theta}_{k+1}^{k+1} \neq \hat{\theta}_j^{k+1}, \forall j \leq k$, claim ii) is true by induction.

**Case 2** $\tilde{\theta}_{k+1} \geq \hat{\theta}_k^k$.

Denote $j$ the smallest integer such that $\hat{\theta}_j^k = \hat{\theta}_{j+1}^k = \cdots = \hat{\theta}_k^k$. Because the boundary is not active between $\theta_{j-1}$ and $\theta_j$, by the pooling condition, we can conclude that $(\hat{\theta}_j^k, \hat{\theta}_{j+1}^k, \cdots, \hat{\theta}_k^k) = arg \min_{\theta_j \geq \cdots \geq \theta_k} \sum_{i=j}^{k} f_i(\theta_i)$ and $\hat{\theta}_j^k = \cdots = \hat{\theta}_k^k = \sum_{i=j}^{k} \tilde{\theta}_i/(k-j+1) \leq \tilde{\theta}_{k+1}$. We claim: $\hat{\theta}_j^m = \hat{\theta}_{j+1}^m = \cdots = \hat{\theta}_{k+1}^m, \forall k \leq m \leq p$. By induction we have already known that $\hat{\theta}_j^m = \hat{\theta}_{j+1}^m = \cdots = \hat{\theta}_k^m$. If $\hat{\theta}_k^m \leq \tilde{\theta}_{k+1}$, then by pooling condition, $f_{k+1}(\theta_{k+1})$ is strictly decreasing on $(0, \hat{\theta}_k^m)$, which forces $\hat{\theta}_{k+1}^m = \hat{\theta}_k^m$. If $\hat{\theta}_k^m \geq \tilde{\theta}_{k+1}$, then $\hat{\theta}_{k+1}^m \geq \tilde{\theta}_{k+1}$ since $f_{k+1}(\theta)$ is unimodal and achieves minimum at $\tilde{\theta}_{k+1}$. Because $\tilde{\theta}_{k+1} \geq \sum_{i=j}^{k} \tilde{\theta}_i/(k-j+1)$, again the pooling condition forces $\hat{\theta}_j^m = \hat{\theta}_{j+1}^m = \cdots = \hat{\theta}_k^m = \sum_{i=j}^{k} \tilde{\theta}_i/(k-j+1)$ and hence $\hat{\theta}_j^m = \hat{\theta}_{j+1}^m = \cdots = \hat{\theta}_k^m = \hat{\theta}_{k+1}^m$. Therefore we proved $\hat{\theta}_j^m = \hat{\theta}_{j+1}^m = \cdots = \hat{\theta}_{k+1}^m, \forall m > k$. Specifically, $\hat{\theta}_j^{k+1} = \hat{\theta}_{j+1}^{k+1} = \cdots = \hat{\theta}_{k+1}^{k+1}$. If $\sum_{i=j}^{k+1} \tilde{\theta}_i/(k-j) \leq \hat{\theta}_{j-1}^k$, still by pooling condition, $\hat{\theta}_j^{k+1} = \cdots = \hat{\theta}_{k+1}^{k+1} = \sum_{i=j}^{k+1} \tilde{\theta}_i/(k-j)$ and consequently $\hat{\theta}_i^{k+1} = \hat{\theta}_i^k, 1 \leq i \leq j-1$. We are done because the solution is exactly PAV$(\tilde{\theta}_1, \cdots, \tilde{\theta}_{k+1})$, which proves claim i) and $\hat{\theta}_j^m = \hat{\theta}_{j+1}^m = \cdots = \hat{\theta}_{k+1}^m, \forall m > k$ implies claim ii). If $\frac{\sum_{i=j}^{k+1} \tilde{\theta}_i}{k-j} > \hat{\theta}_{j-1}^k$, assume $i$ to be the smallest integer such that $\hat{\theta}_i^k = \hat{\theta}_{i+1}^k = \cdots = \hat{\theta}_{j-1}^k$. By similar argument, we can prove that $\hat{\theta}_i^m = \hat{\theta}_{i+1}^m = \cdots = \hat{\theta}_{k+1}^m, \forall m > k$. If $\sum_{t=i}^{k+1} \tilde{\theta}_t/(k-i) < \hat{\theta}_{i-1}^k$, we are done. If not, continue the same argument. □

## 7.2 Proof of Lemma 3.1

Plug in the expression of SURE($\lambda$), we have

$$E[l(\beta_{\hat{\lambda}}, \beta) - SURE(\hat{\lambda})|\beta]$$
$$= E[\frac{1}{p}\sum_{i=1}^{p}\frac{2\hat{\lambda}_i}{\sigma^2+\hat{\lambda}_i}(\tilde{\beta}_i^2 - \tilde{\beta}_i\beta_i - \sigma^2)$$
$$-(\tilde{\beta}_i^2 - \sigma^2 - \beta_i^2)|\beta] = E[\frac{1}{p}\sum_{i=1}^{p}\frac{2\hat{\lambda}_i}{\sigma^2+\hat{\lambda}_i}(\tilde{\beta}_i^2 - \tilde{\beta}_i\beta_i - \sigma^2)|\beta]$$

Take expectation with respect to $\beta$, we get

$$E\{E[l(\beta_{\hat{\lambda}}, \beta) - SURE(\hat{\lambda})|\beta]\} =$$
$$E\{E[\frac{1}{p}\sum_{i=1}^{p}\frac{2\hat{\lambda}_i}{\sigma^2+\hat{\lambda}_i}(\tilde{\beta}_i^2 - \tilde{\beta}_i\beta_i - \sigma^2)|\beta]\}$$

Notice that $\beta_i|\tilde{\beta}_i \sim N(\frac{\sigma_i^2}{\sigma^2+\sigma_i^2}\tilde{\beta}_i, \frac{\sigma^2\sigma_i^2}{\sigma^2+\sigma_i^2})$ and the marginal distribution of $\tilde{\beta}_i$ is $N(0, \sigma^2+\sigma_i^2)$, we change the order of expectation and get:

$$E\{E[l(\beta_{\hat{\lambda}}, \beta) - SURE(\hat{\lambda})|\beta]\} =$$
$$2E[\frac{1}{p}\sum_{i=1}^{p}\frac{\hat{\lambda}_i}{\sigma^2+\hat{\lambda}_i}(\frac{\sigma^2}{\sigma^2+\sigma_i^2}\tilde{\beta}_i^2 - \sigma^2)]$$

where the expectation is with respect to $\tilde{\beta}_i \sim N(0, \sigma^2+\sigma_i^2)$.

$$|E\{E[l(\beta_{\hat{\lambda}}, \beta) - SURE(\hat{\lambda})|\beta]\}| \le$$
$$2\sigma^2 E|\frac{1}{p}\sum_{i=1}^{p}\frac{\hat{\lambda}_i}{\sigma^2+\hat{\lambda}_i}(\frac{\tilde{\beta}_i^2}{\sigma^2+\sigma_i^2} - 1)|$$
$$\le 2\sigma^2 E\{\sup_{1\ge c_1\ge\cdots\ge c_p\ge 0}\frac{1}{p}|\sum_{i=1}^{p}c_i(\frac{\tilde{\beta}_i^2}{\sigma^2+\sigma_i^2} - 1)|\}$$
$$= 2\sigma^2 E\{\sup_{1\ge c_1\ge\cdots\ge c_p\ge 0}\frac{1}{p}|\sum_{i=1}^{p}c_i(Z_i - 1)|\}$$

where $Z_i \sim i.i.d\ \chi^2$. Observe that

$$\sup_{1\ge c_1\ge\cdots\ge c_p\ge 0}|\frac{1}{p}\sum_{i=1}^{p}c_i(Z_i - 1)|$$
$$= \max_{1\le j\le p}|\frac{1}{p}\sum_{i=1}^{j}(Z_i - 1)|$$

which is also used in Lemma 7.2 of Li [1985] and Theorem 3.1 in Xie et al. [2012], we have:

$$|E\{E[l(\beta_{\hat{\lambda}}, \beta) - SURE(\hat{\lambda})|\beta]\}|$$
$$\le 2\sigma^2 E\{\max_{1\le j\le p}|\frac{1}{p}\sum_{i=1}^{j}(Z_i - 1)|\}$$

Let $M_j = \sum_{i=1}^{j}(Z_i - 1)$, then $M_j$ is a martingale. So the $L^2$ maximal inequality implies:

$$E(\max_{1\le j\le p} M_j^2) \le 4E(M_p^2) = 8p$$

$$|E\{E[l(\beta_{\hat{\lambda}}, \beta) - SURE(\hat{\lambda})|\beta]\}|$$
$$\le 2\sigma^2 E\{\max_{1\le j\le p}|\frac{1}{p}\sum_{i=1}^{j}(Z_i - 1)|\}$$
$$\le \frac{2\sigma^2}{p}(E(\max_{1\le j\le p} M_j)^2)^{\frac{1}{2}}$$

Combine the two inequalities, we have

$$|E\{E[l(\beta_{\hat{\lambda}}, \beta) - SURE(\hat{\lambda})|\beta]\}| \le 4\sqrt{\frac{2}{p}}\sigma^2$$

Since the error bound does not depend on $\hat{\lambda}$ and $\sigma_i^2$, the lemma follows. $\square$